\documentclass[a4paper,11pt]{article}
\usepackage{pos}

\usepackage{graphicx,here,epsfig,wrapfig,amssymb,color,amsmath,bm,breqn}

\usepackage[T1]{fontenc} % if needed
\usepackage{multirow}
\usepackage{amsmath}
\usepackage{pifont}
\usepackage{float}
\usepackage{subfig}
\allowdisplaybreaks
\usepackage{soul}
\usepackage[utf8]{inputenc} 
\usepackage{cancel}
\usepackage{multirow,boldline}

\usepackage{lineno} % page numbers
\usepackage[export]{adjustbox}
\usepackage{hyperref} %links in the table of contents
\hypersetup{
    colorlinks,
    citecolor=black,
    filecolor=black,
    linkcolor=black,
    urlcolor=black
}

\renewcommand{\)}{\right)}

\renewcommand{\)}{\right)}

\begin{document}

\title{Flavor-dependent   $b$-profiles  
from Drell-Yan spectra at low transverse momenta}
\ShortTitle{Flavor-dependent  $b$-profiles from Drell-Yan spectra}

\author[a]{M.~Bury}
\author*[b,c,d]{F.~Hautmann}
\author[e]{S.~Leal-Gomez}
\author[f]{I.~Scimemi}
\author[f,g]{A.~Vladimirov}
\author[g]{P.~Zurita}

\affiliation[a]{
Jerzy Haber Institute of Catalysis and Surface Chemistry, Polish Academy
of Sciences, Niezapominajek 8, 30-239 Krak\'ow, Poland}
\affiliation[b]{CERN, Theory Department, CH 1211 Geneva, Switzerland}
\affiliation[c]{Elementaire Deeltjes Fysica, Universiteit Antwerpen, B 2020 Antwerpen, Belgium}
\affiliation[d]{
Theoretical Physics Department, University of Oxford, Oxford OX1 3PU, UK}
\affiliation[e]{
Wien Universit\"at, Faculty of Physics, Boltzmanngasse 5, A-1090 Vienna, Austria}
\affiliation[f]{Dpto. de F\'{i}sica Te\'{o}rica \& IPARCOS, Universidad Complutense de Madrid, E-28040 Madrid, Spain}
\affiliation[g]{Institut f\"ur Theoretische Physik, Universit\"at Regensburg, D-93040 Regensburg, Germany}

\emailAdd{hautmann@thphys.ox.ac.uk}

\abstract{We discuss our recent study of the impact of collinear PDFs  and their 
uncertainties on the determination of transverse momentum dependent (TMD) 
distributions and the description of Drell-Yan (DY) production measurements 
at low transverse momenta. Using QCD factorization and evolution in transverse 
coordinate $b$ space,  this study  takes into account for the first time 
 flavor-dependent non-perturbative $b$-profiles.  It illustrates  
  that collinear PDF uncertainties and non-perturbative TMD flavor dependence  
are both essential to obtain reliable TMD determinations.} 

\FullConference{%
  41st International Conference on High Energy physics - ICHEP2022\\
  6-13 July, 2022\\
  Bologna, Italy
}

\vspace*{-0.6cm} 
\hspace*{11.0cm}  CERN-TH-2022-183 

\maketitle

Transverse momentum spectra in Drell-Yan (DY) vector boson production 
constitute  
one of the key areas of precision strong-interaction physics at hadron 
colliders. 
The QCD factorization formula for the DY differential cross section in the 
vector boson transverse momentum $q_T$~\cite{Collins:2011zzd} 
 at $q_T \ll Q$, where $Q$ is the vector boson invariant mass,   
enables one to explore perturbative dynamics, embodied in logarithmic 
resummations to all orders in the strong coupling $\alpha_s$, as well as  
non-perturbative dynamics, embodied in  transverse momentum 
dependent (TMD) parton distributions~\cite{Angeles-Martinez:2015sea}. 
Phenomenological studies of DY $q_T$ spectra and their impact on 
precision electroweak measurements~\cite{Lhcew:2022pt} are 
 carried out 
 at present 
with a variety of computational approaches, from 
the classic {\sc Resbos} platform~\cite{Isaacson:2022rts} to 
the recent tools~\cite{Camarda:2022qdg,Coradeschi:2017zzw,Chen:2022cgv,BermudezMartinez:2019anj,Ebert:2020dfc,Becher:2020ugp} 
to the global TMD fits~\cite{Bacchetta:2019sam,Scimemi:2019cmh}. 

According to the factorization formula,  the DY $q_T$ differential cross 
section may be written schematically, up  to power corrections in $q_T/Q$ and  
$\Lambda_{\rm{QCD}} / Q$, as  
\begin{equation} 
\label{eq-partonfact} 
{{d \sigma} \over { d q_T^2} } = \sum_{i,j} \int d^2 b \ e^{i b \cdot q_T} \sigma^{(0)}_{i j} f_{i }(x_1,b;\mu,\zeta_1)f_{j}(x_2,b;\mu,\zeta_2) \;\; ,    
\end{equation} 
where 
the indices $i , j$ run over quark and antiquark flavors, 
$b$ is the transverse distance Fourier conjugate to $q_T$,  
$ \sigma^{(0)}_{i j}$ are perturbatively calculable hard-scattering 
functions, and $f_{i}$ and $f_{j}$ are TMD parton distributions, 
 evolving with the mass and rapidity scales $\mu$ and $\zeta$ according to 
well-prescribed  evolution equations.  
Eq.~(\ref{eq-partonfact}) is the basis for 
 the phenomenological program advocated 
 in~\cite{Hautmann:2014kza}, which 
consists in determining TMD distributions,  
via the factorization and evolution  framework, 
 from  fits to experimental data, in a manner  
analogous to  (but independent of) the case of ordinary, collinear   
    parton distribution functions (PDFs).

In practice,  many current 
studies employ, besides the  factorization and evolution 
 formulas, the  operator product   expansion  (OPE) 
 of   TMD distributions in terms of 
 PDFs.  This is valid at small distances 
 $b$,  up to power corrections in $b \Lambda_{\rm{QCD}} $, and reads 
\begin{equation}
\label{eq-ope}
f_{i}(x,b;\mu,\zeta)=\sum_{j}\int_{x}^1 (dy/y) C_{i j}\(y,b;\mu,\zeta\)
q_{j} (x/y,\mu)+{\cal O} (b \Lambda_{\rm{QCD}})^2 \;\; ,
\end{equation}
where  $q_j(x,\mu)$ are the PDFs, and $C_{i j}$ are 
perturbatively calculable coefficient functions. 

In particular, 
to carry out TMD fits 
an ansatz is made for the non-perturbative (NP) TMD distributions at large 
$b$, where the power corrections to Eq.~(\ref{eq-ope})  become sizeable,  
of the form 
\begin{equation}
\label{eq-opeansatz}
f_{i}(x,b;\mu,\zeta)=\sum_{j}\int_{x}^1 (dy/y) C_{i j}\(y,b;\mu,\zeta\)
q_{j}(x/y,\mu) 
 f_{\text{NP}}^i(x,b) \; . 
\end{equation}
Here the distributions  $f_{\text{NP}}^i$ 
behave as 
 $f_{\text{NP}}^i(x,b)\sim 1  + 
{\cal O} (b \Lambda_{\rm{QCD}})^2$ for $ b \to 0$, thus matching 
   the OPE (\ref{eq-ope}), 
 and are to be determined from experimental data,   
 once a choice is made for the collinear PDF in Eq.~(\ref{eq-opeansatz}).

\begin{figure}[t]
\centering
\includegraphics[width=0.99\textwidth]{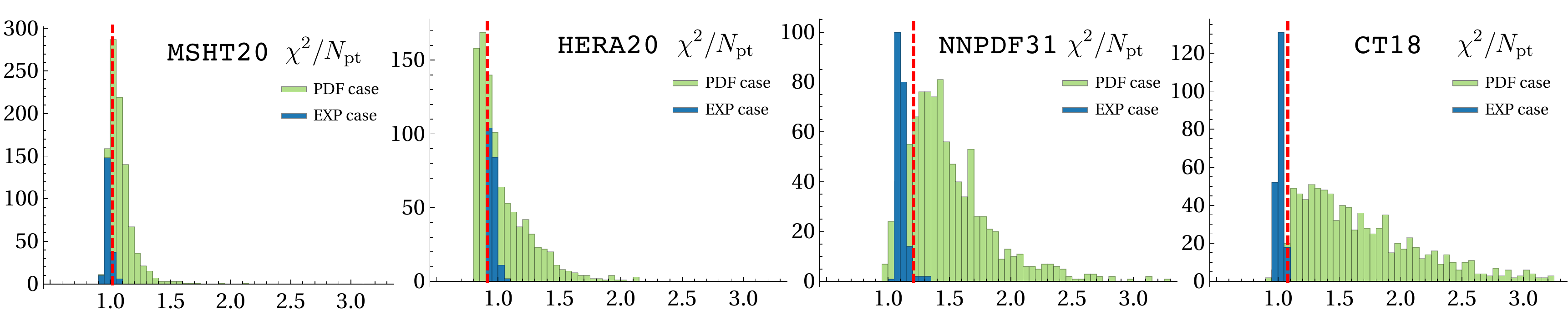}
\caption{\label{fig:distribofchi2}
Distribution of $\chi^2$ values for  {\bf PDF} and {\bf EXP} 
cases. The position of the final $\chi^2$ is indicated 
by the red line.}
\end{figure}

While earlier studies have assumed flavor-independent $f_{\text{NP}}$, 
 Ref.~\cite{Bury:2022czx}  explores for the first time the flavor dependence in 
the  non-perturbative $f_{\text{NP}}$ $b$-profiles. The implications of 
propagating  uncertainties from the PDF $q_j$ in 
the ansatz (\ref{eq-opeansatz}) to the extracted NP TMD  
$f_{\text{NP}}$ via Eq.~(\ref{eq-partonfact}) have also 
 not been addressed in the literature before, and are explored for the first time 
 in Ref.~\cite{Bury:2022czx}. 

For the study~\cite{Bury:2022czx} we use the implementation of TMD 
factorization and evolution at the next-to-next-to-leading-logarithmic 
level as in~\cite{Hautmann:2020cyp}  
(NNLL$^\prime$ according to 
the terminology adopted in~\cite{Lhcew:2022pt}); we 
 perform the analysis 
 for the   PDF sets  
 \texttt{MSHT20}~\cite{Bailey:2020ooq}, 
 \texttt{CT18}~\cite{Hou:2019efy}, 
   \texttt{NNPDF3.1}~\cite{Ball:2017nwa},  
\texttt{HERA20}~\cite{Abramowicz:2015mha}, taken 
as representatives of different methodological approaches at 
next-to-next-to-leading order (NNLO); we  
 carry out  fits of experimental data for 
low-$q_T$ DY production at  
fixed-target~\cite{Ito:1980ev,Moreno:1990sf,McGaughey:1994dx},  
   RHIC~\cite{PHENIX:2018dwt}, 
   Tevatron~\cite{Affolder:1999jh,Abbott:1999wk}  and 
   LHC~\cite{Aad:2014xaa,Chatrchyan:2011wt,Aaij:2015gna} 
   experiments. 
   Unlike previous TMD determinations 
   in which TMD uncertainties are 
   obtained from experimental uncertainties, 
   in the analysis~\cite{Bury:2022czx} TMD uncertainties 
   result from both experiment (EXP) and PDF.  The two sources of 
   uncertainties are taken into account through a Bayesian approach, 
   in which PDFs are represented as Monte Carlo ensembles, and 
   uncertainties 
   are obtained by fitting each member of the input ensemble.  
   The bootstrap method~\cite{Efron:1979bxm} is 
   used to address the issue of combining EXP and PDF sources. 
The key findings are as follows.

\begin{table}
\footnotesize
\begin{center}
\renewcommand{\arraystretch}{0.94}
\begin{tabular}{|l | c V{4} c V{4} c V{4} c V{4} c V{4}}
\cline{3-6}
\multicolumn{2}{c V{4}}{ } & 
\textbf{MSHT20}  &
\textbf{HERA20}  &
\textbf{NNPDF31}  &
\textbf{CT18}  \\
\hline 
Data set & $N_{pt}$ & 
$\chi^2/N_{pt}$ & 
$\chi^2/N_{pt}$ & 
$\chi^2/N_{pt}$ & 
$\chi^2/N_{pt}$  
\\\hline
CDF run1          &   33 &      0.78 & 0.61	& 0.72 	& 0.75 \\
CDF run2          &   39 &      1.70 & 1.42	& 1.68 	& 1.79 \\
D0 run1           &   16 &      0.71 & 0.81 & 0.79  & 0.79 \\
D0 run2           &    8 &      1.95 & 1.39	& 1.92  & 2.00 \\
D0 run2 ($\mu$)   &    3 &      0.50 & 0.59	& 0.55  & 0.52 \\
\hline
ATLAS 7TeV 0.0<|y|<1.0      &    5 &      4.06 & 1.94 & 2.12 & 4.21\\
ATLAS 7TeV 1.0<|y|<2.0      &    5 &      7.78 & 4.83 & 4.52 & 6.12\\
ATLAS 7TeV 2.0<|y|<2.4      &    5 &      2.57 & 2.18 & 3.65 & 2.39\\
ATLAS 8TeV 0.0<|y|<0.4      &    5 &      2.98 & 3.66 & 2.12 & 3.23\\
ATLAS 8TeV 0.4<|y|<0.8      &    5 &      2.00 & 1.53 & 4.52 & 3.21\\
ATLAS 8TeV 0.8<|y|<1.2      &    5 &      1.00 & 0.50 & 2.75 & 1.89\\
ATLAS 8TeV 1.2<|y|<1.6      &    5 &      2.25 & 1.61 & 2.49 & 2.72\\
ATLAS 8TeV 1.6<|y|<2.0      &    5 &      1.92 & 1.68 &	2.86 & 1.96\\
ATLAS 8TeV 2.0<|y|<2.4      &    5 &      1.35 & 1.14 & 1.47 & 1.06\\
ATLAS 8TeV 46<Q<66GeV      &    3 &       0.59 & 1.86 & 0.23 & 0.05\\
ATLAS 8TeV 116<Q<150GeV    &    7 &       0.61 & 1.03 & 0.85 & 0.70\\
\hline
CMS 7TeV          &    8 &      1.22 & 1.19 & 1.30 & 1.25\\
CMS 8TeV          &    8 &      0.78 & 0.77 & 0.75 & 0.78\\
CMS 13TeV 0.0<|y|<0.4 &    8 &       3.52& 1.93 & 2.13& 3.73\\
CMS 13TeV 0.4<|y|<0.8 &    8 &       1.06& 0.53 & 0.71& 1.65\\
CMS 13TeV 0.8<|y|<1.2 &    10 &      0.48& 0.14 & 0.33& 0.88\\
CMS 13TeV 1.2<|y|<1.6 &    11 &      0.62& 0.33 & 0.47& 0.86\\
CMS 13TeV 1.6<|y|<2.4 &    13 &      0.46& 0.32 & 0.39& 0.57\\
\hline
LHCb 7TeV         &    8 &      1.79 & 1.00 & 1.62 & 1.16\\
LHCb 8TeV         &    7 &      1.38 & 1.29 & 1.63 & 0.83\\
LHCb 13TeV        &    9 &      1.28 & 0.84 & 1.07 & 0.93\\
\hline
PHE200             &    3 &      0.29 &	0.42 & 0.38 & 0.29 \\
E288-200           &   43 &      0.43 & 0.36 & 0.57 & 0.43 \\
E288-300 $Q<9$GeV  &   43 &      0.77 & 0.56 & 0.89 & 0.55 \\
E288-300 $Q>11$GeV &   10 &      0.29 & 0.37 & 0.45 & 0.44 \\
E288-400 $Q<9$GeV  &   34 &      2.19 & 1.15 & 1.49 & 1.34 \\
E288-400 $Q>11$GeV &   42 &      0.25 & 0.61 & 0.44 & 0.40 \\
E772               &   35 &      1.14 & 1.37 & 1.79 & 1.11 \\
E605     $Q<9$GeV  &   21 &      0.52 & 0.47 & 0.47 & 0.61 \\
E605     $Q>11$GeV &   32 &      0.47 & 0.73 & 1.34 & 0.52\\
\hline\hline
\textbf{Total} & \textbf{507} & \textbf{1.12} & \textbf{0.91} & \textbf{1.21} & \textbf{1.08} \\
\hline
\end{tabular}
\caption{\label{chi2table} The $\chi^2$ values for the central replica over 
the TMD data set  for different PDF input.}
\end{center}
\end{table}

\begin{figure}
\centering
\includegraphics[width=\textwidth]{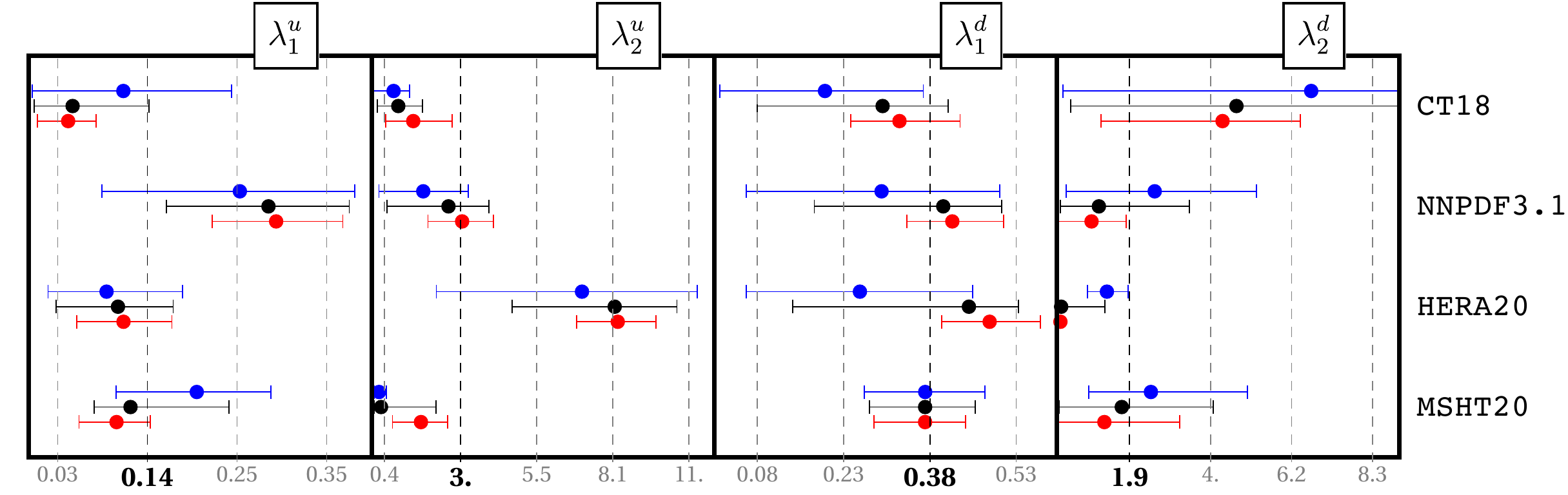}
\caption{\label{fig:lambda1e2}
Fitted NP TMD parameters:  (blue) value from the 
fit of the {\bf PDF} case;  (red) value from the fit of the {\bf EXP} case; 
(black) final result. 
}
\end{figure}

\begin{figure}
\centering
\includegraphics[width=0.85\textwidth]{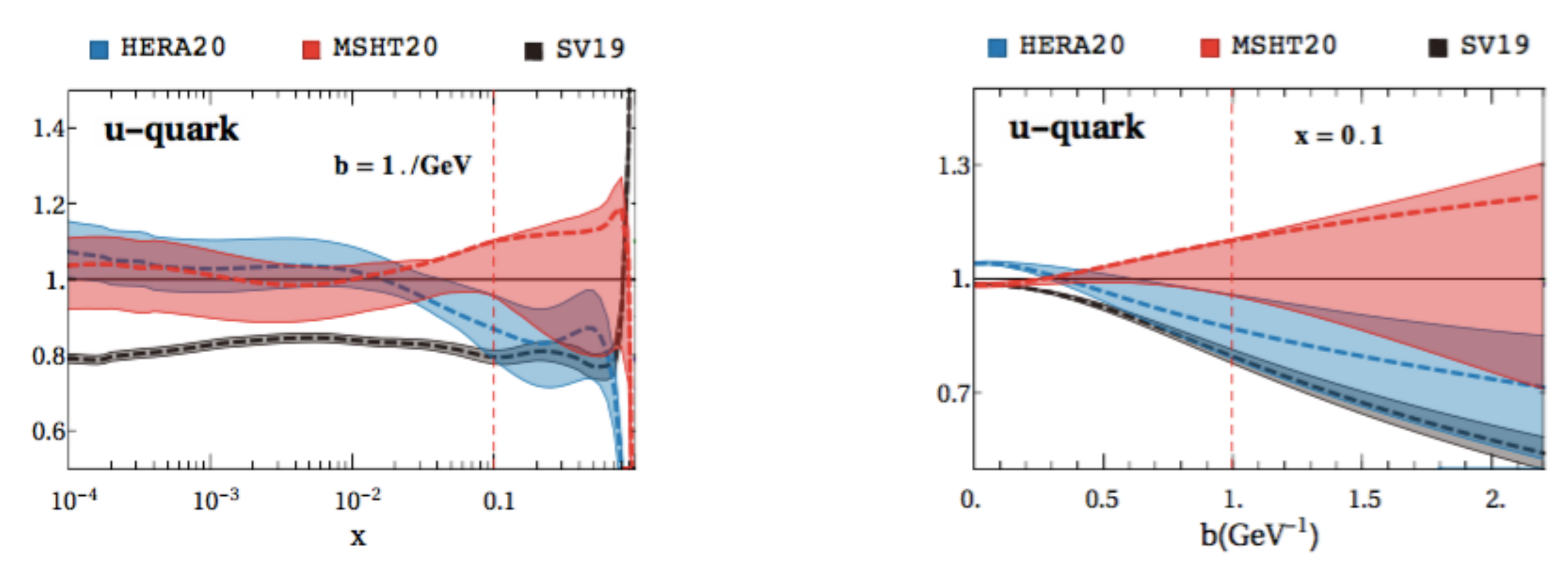}
\caption{\label{fig:comp-vs-b-vs-x}
TMD uncertainty bands  as functions of $x$ and $b$. 
The plot is weighted with the central TMD value 
averaged over different PDF sets. For comparison  the result 
SV19~\cite{Scimemi:2019cmh} is also shown.  
}
\end{figure}

(i)  An overall reasonable description of the data is obtained for 
all PDF sets, with total $\chi^2$ values of the fits and individual 
$\chi^2$ values for each experiment given in Table~\ref{chi2table}. 
The $\chi^2$ distribution among PDF and EXP replicas is shown in 
Fig.~\ref{fig:distribofchi2}. 
In all cases the PDF uncertainty is found to be larger than the EXP 
uncertainty. 
If flavor-independent $b$-profiles are assumed, as 
e.g.~in~\cite{Scimemi:2019cmh}, 
the spread of $\chi^2$ values 
is found to be much larger than in  Fig.~\ref{fig:distribofchi2}. 

(ii) Flavor-dependent TMD parameters in 
$f_{\text{NP}}^i$  are extracted from the fits. 
An example is given in Fig.~\ref{fig:lambda1e2}, where 
 $f_{\text{NP}}^i (x , b)$ are parameterized through 
 a combination of exponential and gaussian $b$-profiles
with  flavor-dependent $\lambda$ coefficients (while the NP 
contribution to the evolution kernel is 
parameterized through a linear behavior at large 
$b$ ---  see~\cite{Hautmann:2020cyp} for discussion of alternative 
parameterizations, including a constant large-$b$ behavior 
 in the spirit of the picture~\cite{Hautmann:2007cx}). 

(iii) With the analysis 
including collinear PDF effects and flavor-dependent $b$-profiles, 
TMD error bands are 
significantly increased compared to earlier OPE-based fits.  
The bands are illustrated in Fig.~\ref{fig:comp-vs-b-vs-x} versus $x$ and 
versus $b$. A 
comparison with the bands from the earlier fit~\cite{Scimemi:2019cmh}
is also shown. 

We conclude by noting that variations in PDF replicas are found to 
affect not only the normalization  but also the $q_T$ shape 
of TMD predictions. This causes the observed spread in 
$\chi^2$ values among replicas. It is due to the OPE coupling the 
$b$ and $x$ dependences. 
Also, TMD uncertainty 
contributions  from PDFs are  
comparable in size to  theoretical uncertainties from perturbative 
scale variations for low-$q_T$ DY observables, pointing to the 
relevance of joined TMD and PDF fits.

%\vskip 0.2 cm 

\noindent {\bf Acknowledgments}. We thank the ICHEP2022 organizers and 
convenors for the invitation and the very interesting conference.

\end{document}